\documentclass{article}
\usepackage{amsmath}
\usepackage{graphicx}
\usepackage{amsfonts}
\usepackage{amssymb}

\begin{document}

\markboth{de Melo and Resende}
{Galactic Nonlinear Dynamic Model}

%
%

\title{Galactic Nonlinear Dynamic Model\footnote{In celebration of the 60th anniversary of Prof. B. M. Pimentel.}
}

\author{C. A. M. de Melo$^{1,2}$ and S. T. Resende$^2$\\
$^1$Instituto de F\'{\i}sica Te\'{o}rica, UNESP - S\~{a}o Paulo State University.\\
Rua Pamplona 145, CEP 01405-900, S\~{a}o Paulo, SP, Brazil.\\
$^2$Universidade Vale do Rio Verde de Tr\^{e}s Cora\c{c}\~{o}es,\\
Av. Castelo Branco, 82 - Ch\'{a}cara das Rosas, P.O. Box 3050,\\
CEP 37410-000, Tr\^{e}s Cora\c{c}\~{o}es, MG, Brazil.\\
cassius.anderson@gmail.com}

\maketitle

\begin{abstract}
We develop a model for spiral galaxies based on a nonlinear
realization of the Newtonian dynamics starting from the momentum and mass
conservations in the phase space. The radial solution exhibits a rotation
curve in qualitative accordance with the observational data.
\end{abstract}



\section{Introduction}

Galactic rotation curves are one of the most important evidences favoring the
dark matter scenario in astrophysics. Actually dark matter was proposed by
Zwicky \cite{Zwicky} in order to accommodate observed velocities in galaxies
and galaxy clusters. Since then others explanations was proposed as
modifications of Newtonian gravity \cite{MOND} or modifications in General
Relativity \cite{ModGR}. All these approaches have in common, from the
modeling point of view, just one element: non-linearity. So, is natural asks
the question if non-linear effects of some sort are responsible for the
discrepancy observed in galaxies and galaxy cluster dynamics. Recently, in the
context of General Relativity, some authors have investigated this point
\cite{Maia}.

Here, we will propose an alternative paradigm where Newtonian gravity is
maintained on its solid basements but a non-linear model of a galaxy is build.
We solve the non-linear equations in two simplified cases and calculate the
resulting galaxy rotation curves, showing qualitatively that flat curves can
be obtained in a given region of the parameter space.

\section{The General Model}

We focus on spiral galaxies described as a fluid with mass distribution in the
phase space, $\Psi\left(  \mathbf{x},\mathbf{v},t\right)  $ which is related
to the ordinary mass distribution by%
\[
\rho\left(  \mathbf{x},t\right)  =m_{\ast}\int d\mathbf{v~}\Psi\left(
\mathbf{x},\mathbf{v},t\right)
\]
where $m_{\ast}$ the mass for a typical star. the dynamics is governed by the
Poisson equation coming from the second Newton law,%
\[
\mathbf{\nabla}^{2}V\left(  \mathbf{x},t\right)  =4\pi G\rho\left(
\mathbf{x},t\right)
\]
but restricted to the mass conservation,%
\[
\frac{d\Psi}{dt}=0\rightarrow\frac{\partial\Psi}{\partial t}+\frac
{\partial\Psi}{\partial\mathbf{x}}\cdot\mathbf{\dot{x}}+\frac{\partial\Psi
}{\partial\mathbf{v}}\cdot\mathbf{\dot{v}}=0
\]

Using a Hamiltonian description, $\mathbf{\dot{x}=v}$\textbf{, }%
$\mathbf{\dot{v}}=-\mathbf{\nabla}V$, we obtain,%
\begin{gather*}
\frac{\partial\Psi}{\partial t}+\mathbf{v}\cdot\mathbf{\nabla}\Psi
-\mathbf{\nabla}V\cdot\frac{\partial\Psi}{\partial\mathbf{v}}=0\\
\mathbf{\nabla}^{2}V\left(  \mathbf{x},t\right)  =4\pi G\int d\mathbf{v~}%
\Psi\left(  \mathbf{x},\mathbf{v},t\right)
\end{gather*}

In general, this is a set of integro-differential non-linear coupled equations.

\subsection{Axial Symmetry}

Let us restrict the model only to the disc of the galaxy in the static regime.
So, using the axial symmetry of this system, the mass conservation equation
become:%
\[
\dot{r}\frac{\partial\Psi}{\partial r}+\dot{\phi}\frac{\partial\Psi}%
{\partial\phi}+\dot{z}\frac{\partial\Psi}{\partial z}-\left(  \frac{\partial
V}{\partial r}\frac{\partial\Psi}{\partial\dot{r}}+\frac{\partial V}%
{\partial\phi}\frac{\partial\Psi}{\partial\dot{\phi}}+\frac{\partial
V}{\partial z}\frac{\partial\Psi}{\partial\dot{z}}\right)  =0
\]

This equation can be solved using the Method of Characteristics. Assuming a
constant angular velocity, this is equivalent to a set of ordinary
differential equations,%
\[
\frac{dr}{\dot{r}}=\frac{d\dot{r}}{-\frac{\partial V}{\partial r}}=\frac
{dz}{\dot{z}}=\frac{d\dot{z}}{-\frac{\partial V}{\partial z}}~,\;\frac{d\phi
}{\dot{\phi}}=0~,\;\frac{d\Psi}{dt}=0
\]
whose uniparametric family of solutions is%
\[
\Psi=\Psi\left(  E\right)  ~,\;E=\frac{1}{2}\left(  \dot{r}^{2}+r^{2}\dot
{\phi}^{2}+\dot{z}^{2}\right)  +V\left(  r,z\right)
\]

Experimental data shows that the mass distribution is Gaussian in the observed
velocities, therefore is natural to choose the mass distribution $\Psi$ to be
a Boltzmann distribution in the total energy:%
\[
\Psi\left(  E\right)  =\Psi_{0}e^{-\beta E}=\Psi_{0}\exp\left(  -\beta\left[
\frac{1}{2}\left(  \dot{r}^{2}+r^{2}\dot{\phi}^{2}+\dot{z}^{2}\right)
+V\left(  r,z\right)  \right]  \right)
\]

The integration over the velocities can be done, in order to obtain the
functional dependence of the mass distribution over the gravitational
potential%
\[
\rho\left(  r,z\right)  =\int d\dot{r}d\dot{\phi}d\dot{z}r\Psi\left(
E\right)  =\Psi_{0}\left(  \frac{2\pi}{\beta}\right)  ^{3/2}\exp\left(  -\beta
V\left(  r,z\right)  \right)  =\rho_{0}e^{-\beta V\left(  r,z\right)  }%
\]

It means that the Poisson equation now is a non-linear partial differential
equation, given by%
\[
\frac{\partial^{2}V}{\partial r^{2}}+\frac{1}{r}\frac{\partial V}{\partial
r}+\frac{\partial^{2}V}{\partial z^{2}}=4\pi G\rho_{0}e^{-\beta V\left(
r,z\right)  }%
\]

\section{Variation in the Height}

Let us to take an over simplified case, assuming that the field vary only with
the height to the plane of the disc,%
\[
\frac{d^{2}V}{dz^{2}}=4\pi G\rho\left(  z\right)  =4\pi G\rho_{0}e^{-\beta
V\left(  z\right)  }%
\]
Using an integration factor $\frac{dV}{dz}$ and the boundary conditions
$V\left(  0\right)  =\frac{dV\left(  0\right)  }{dz}=0$ this equation can be
directly integrated,%
\[
V\left(  z\right)  =\frac{2}{\beta}\ln\cosh\left(  \frac{z}{z_{0}}\right)
~,\;z_{0}\equiv\left(  2\pi G\rho_{0}\beta\right)  ^{-1/2}%
\]

The density profile in this case is%
\[
\rho\left(  z\right)  =\rho_{0}e^{-\beta V}=\frac{\rho_{0}}{\cosh^{2}\left(
\frac{z}{z_{0}}\right)  }%
\]

So far, it is a reasonable model, since the disc predicted is thin.

\section{Radial Variation}

Let us to take another over simplified model assuming variation only in the
radial direction,%
\[
\frac{d^{2}V}{dr^{2}}+\frac{1}{r}\frac{dV}{dr}=4\pi G\rho_{0}e^{-\beta
V\left(  r\right)  }%
\]

Multiplying both sides by $r$ and integrating, we find%
\[
r\frac{dV}{dr}-r_{0}c=4\pi G\rho_{0}\int_{r_{0}}^{r}\tilde{r}e^{-\beta
V\left(  \tilde{r}\right)  }d\tilde{r}%
\]
where $r_{0}$ is the radius of the galaxy core, and $c\equiv\frac{dV\left(
r_{0}\right)  }{dr}$. Performing a second integration, we arrive in a Volterra
second order integral equation:%
\[
V\left(  r\right)  =V_{0}+r_{0}c\ln\left(  \frac{r}{r_{0}}\right)  +4\pi
G\rho_{0}\int_{r_{0}}^{r}d\bar{r}\int_{r_{0}}^{\bar{r}}\frac{\tilde{r}}%
{\bar{r}}e^{-\beta V\left(  \tilde{r}\right)  }d\tilde{r}%
\]

To solve it, we apply the Piccard's method of successive approximations:%
\[
V^{\left(  n+1\right)  }\left(  r\right)  =V^{\left(  0\right)  }\left(
r\right)  +4\pi G\rho_{0}\int_{r_{0}}^{r}d\bar{r}\int_{r_{0}}^{\bar{r}}%
\frac{\tilde{r}}{\bar{r}}e^{-\beta V^{\left(  n\right)  }\left(  \tilde
{r}\right)  }d\tilde{r}%
\]%
\[
V^{\left(  0\right)  }\left(  r\right)  =V_{0}+r_{0}c\ln\left(  \frac{r}%
{r_{0}}\right)
\]
Therefore, the first order iterative solution is%
\begin{align*}
V^{\left(  1\right)  }\left(  r\right)   &  =V_{0}-4\pi G\rho_{0}e^{-\beta
V_{0}}\left(  \frac{r_{0}}{2-\beta r_{0}c}\right)  ^{2}+\left(  r_{0}%
c-\frac{4\pi r_{0}^{2}G\rho_{0}e^{-\beta V_{0}}}{2-\beta r_{0}c}\right)
\ln\left(  \frac{r}{r_{0}}\right)  +\\
&  +4\pi G\rho_{0}e^{-\beta V_{0}}\left(  \frac{r}{2-\beta r_{0}c}\right)
^{2}\left(  \frac{r_{0}}{r}\right)  ^{\beta r_{0}c}%
\end{align*}

\section{Galaxy Rotation Curve}

Assuming a virial balance, $\frac{\upsilon^{2}}{r}=\frac{dV}{dr}%
\rightarrow\upsilon=\sqrt{r\frac{dV}{dr}}$, the galaxy rotation curve in this
model is $\left(  \bar{\beta}\equiv\beta r_{0}c~,\;\left(  r_{G}\right)
^{-2}\equiv4\pi G\rho_{0}~,\;\lambda\equiv\frac{r}{r_{0}}~,\;a\equiv
\frac{r_{0}}{r_{G}}\right)  $%
\[
\upsilon^{\left(  1\right)  }\left(  \lambda\right)  =\sqrt{\frac{\bar{\beta}%
}{\beta}+\left(  \frac{a^{2}}{2-\bar{\beta}}\right)  \left(  \lambda
^{2-\bar{\beta}}-1\right)  }%
\]

\begin{figure}[th]
\begin{center}
\includegraphics[width=13cm]{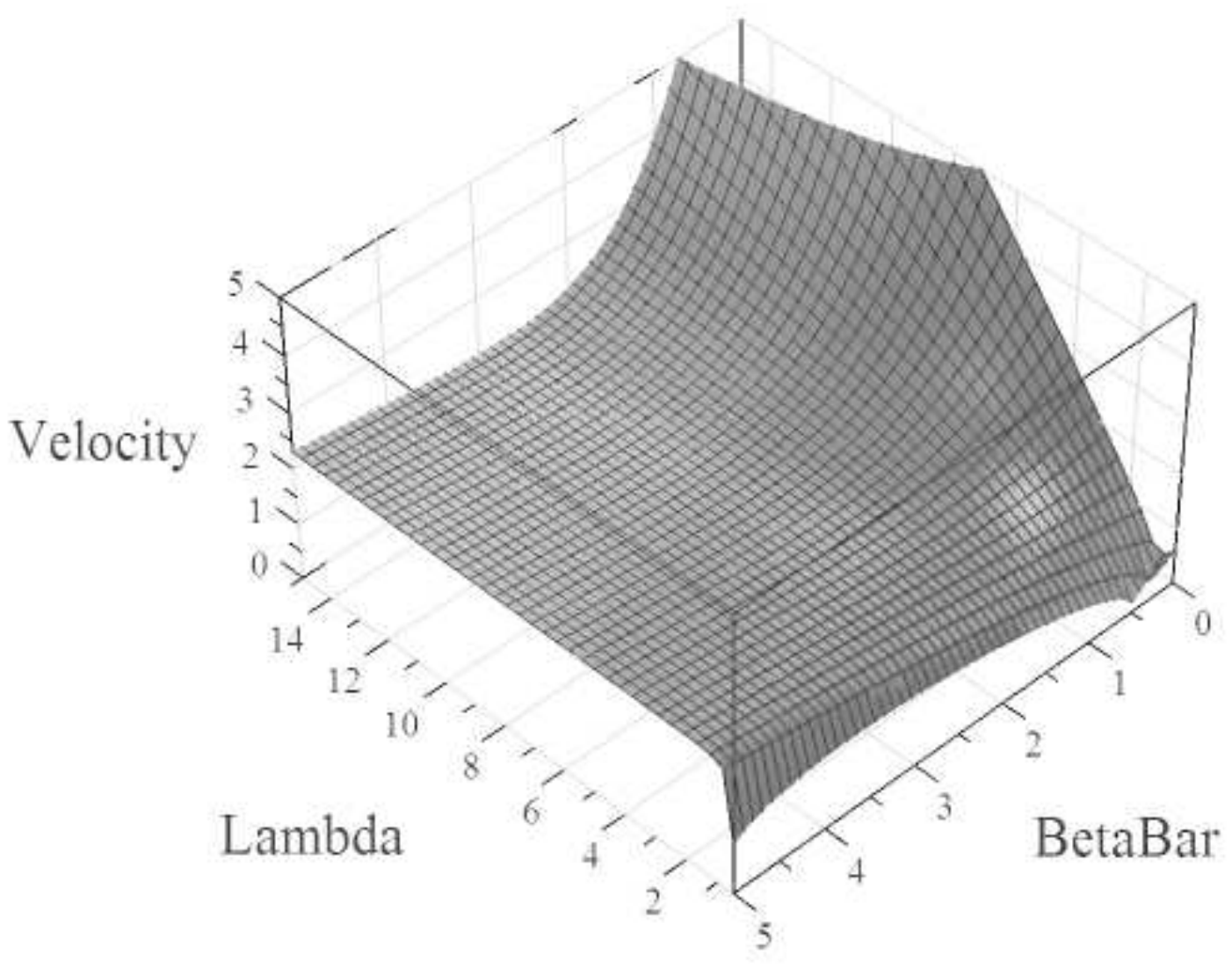}
\end{center}

\vspace*{8pt}
\caption{Parametric rotation curve. Velocity is given in units of $1/\beta^{1/2}$ and $\lambda$ is the distance to center in adimensional units. $\bar{\beta}$ is the parameter dictated by the boundary condition at the core. \label{f1}}
\end{figure}

In our illustrative example, $a=\beta=1$, we have the behavior illustrated in Fig.~\ref{f1}.

So, if $\bar{\beta}>2$,%
\[
\lim_{\lambda\rightarrow\infty}\upsilon^{\left(  1\right)  }\left(
\lambda\right)  =\sqrt{\frac{\bar{\beta}}{\beta}-\left(  \frac{a^{2}}%
{2-\bar{\beta}}\right)  }%
\]
which is constant.

\section{Final Remarks}

A non-linear Newtonian model for the galaxy disk was constructed by
introducing a solution of the mass conservation on the phase space.

We illustrate by a direct example that the non-linear character of the
gravitational field is a key feature to understand galaxy rotation curves,
even in the Newtonian case.


\begin{thebibliography}{9}                                                                                                %
\bibitem {Zwicky}F. Zwicky, \textit{Helv. Phys. Acta }\textbf{6} (1933) 110.

\bibitem {MOND}J. R. Bownstein and J. W. Moffat, \textit{Astrophys. J.}
\textbf{636} (2006) ??; M. Milgrom, \textit{Astrophys. J.} \textbf{270} (1983)
365, \textit{ibid.} 371, \textit{ibid.} 384.

\bibitem {ModGR}S. Fay, \textit{Astron. Astrophys.} \textbf{413} (2004) 799;
S. Behar and M. Carmeli, \textit{Int. J. Mod. Phys.} \textbf{39} (2000) 1397;
S. Capozziello \textit{et al., Phys. Lett.} \textbf{A326} (2004) 292; D. N.
Vollick, \textit{Gen. Rel. \& Grav.} \textbf{34} (2002) 471; K. Ichiki
\textit{et al., Phys. Rev. }\textbf{D66} (2002) 023514; R. R. Cuzinatto, C. A. M. de Melo, L. G. Medeiros and P. J.
Pompeia, \textit{Eur. Phys. J. }\textbf{C53} (2008) 98.

\bibitem {Maia}M. D. Maia, A. J. S. Capistrano and D. M\"{u}ller,
\textit{astro-ph/0605688}.
\end{thebibliography}
\end{document}